\documentstyle[epsf]{mn}
\def\etal{{\rm et al.~}}

\def\sub{{substructures}}
\def\apim{{$\Pi^{(m)}$}}
\def\api2{{$\Pi^{(2)}$}}
\def\api3{{$\Pi^{(3)}$}}
\def\api4{{$\Pi^{(4)}$}}
\def\h80{{$h_{80}^{-1}$}}

\title[Evolution of cluster morphology]
{Evolution of Cluster Morphology: X--ray data $vs$ CDM models}

\author[R.Valdarnini et al.]
{Riccardo Valdarnini$^{1}$, Simona Ghizzardi$^{2,3}$ and
Silvio Bonometto$^{2,3}$\\ \\
$^1$SISSA -- International School for Advanced Studies,
    Via Beirut 2/4, Trieste, Italy\\  
$^2$Department of Physics of the University,
    Via Celoria 16, Milano, Italy\\
$^3$INFN -- Sezione di Milano }

\date{Accepted 1997 ** **. Received 1997 ** **; in original form 1997 ** **}

\begin{document}

\def\ltsima{$\; \buildrel < \over \sim \;$}
\def\simlt{\lower.5ex\hbox{\ltsima}}
\def\gtsima{$\; \buildrel > \over \sim \;$}
\def\simgt{\lower.5ex\hbox{\gtsima}}

\maketitle

\begin{abstract}

We compare the evolution of morphology in X--ray cluster data
and in clusters obtained in a simulation of flat CDM,
normalized to the observed cluster abundance. We find that
the evolution rate in model clusters is significantly higher
than in data clusters. The results is stricking, as our cluster data
are all just for $z \simlt 0.2$, but is not contrary to expectations,
as is known that the CDM spectrum is too flat, around the cluster scale,
and therefore induces the presence of too many substructures.
The cluster simulations were run using a TREESPH code and
therefore include hydrodynamical effects. The test we performed,
which turns out to be very sensitive,
starts from the so--called power ratios introduced by Buote and Tsai, and
makes use of the 2--dimensional generalization of the Kolmogorov--Smirnov
test. The discrepancy between data and model is however directly
visible through linear regression techniques and
using simpler statistical tests.

\end{abstract}


\footnotetext[2]{E-mail: valda@sissa.it;\\
ghizzardi@mi.infn.it;\\
bonometto@mi.infn.it}

\section{Introduction}

The principal aim of this work is to study the evolution of morphology 
in hydrodynamical simulations of galaxy clusters, for 
flat CDM cosmological models, normalized in order to reproduce the observed 
cluster abundance. We also compare its trend with observational 
X--ray data and find a substantial discrepancy on the rate of evolution 
of substructures, which seem to evolve faster in models than in data. 

This result is not surprising, as the slope of 
CDM spectrum, around the cluster scale, is known to be too flat. 
E.g., the {\sl extra power} parameter
$\Gamma = 7.13 \cdot 10^{-3}\left( \sigma_8 / \sigma_{25} \right)^{10/3}$
($\sigma_{8,25}$ are mass variances on $8,25 ~ h^{-1}$Mpc scales; $h=H/100~
{\rm km}\, {\rm s}^{-1}{\rm Mpc}^{-1}$) is significantly lower in 
data than in model predictions: for APM galaxies, Peacock and Dodds (1995) 
found $\Gamma =0.23 \pm 0.04$; for the Abell/ACO sample, Borgani \etal 
(1997) found $\Gamma$ in the interval 0.18--0.25; on the contrary CDM 
models predict $\Gamma \simeq 0.4\, $.

What is however stricking is the sensitivity that the tests we performed on
hydrodynamical simulations, seem to have in appreciating the spectral slope.
High $\Gamma$ values imply that too many \sub ~intervene in the dynamical
growth of fluctuations, on cluster scales. An excess evolution can be therefore
expected. But available data cover only a narrow redshift interval ($0 < z <
0.2$) and simulation outputs shall be used on the same interval, reproducing
data distribution. The point is then that different rates seem already visible
on such redshifts, which concern a fairly recent cosmic evolution. 

The tests we use start from the a statistical tool introduced by 
Buote \& Tsai (1995), who defined the so--called power ratios \apim,
which essentially derive from a multipole expansion of X--ray surface 
brightness. In sec. 2 more details on \apim definition will be reported.
Much  work on \apim was already done, both in order to derive them from 
data and to compare them with simulations 
(Buote \& Tsai 1995, Buote \& Tsai 1996, Tsai \& Buote 1996,
Buote \& Xu 1997). In a recent work, Buote e Xu
(1997) gave some plots where the evolution of cluster morphology
in data and CDM simulations are compared; similar plots are given also by us.

Here, however, we present two essential improvements: (i) Instead of 
using pure CDM N--body simulations, we use the outputs of a TREESPH code, 
therefore including hydrodynamical effects. A large number of massive clusters
are treated in this way and hydrodynamical effects are found
to be important in shaping the distribution of the X--ray emitting
baryon component. This will be shown in some plots where DM and baryon
distributions are compared; but there is also a back effect of the
different baryon distributions on the evolution of the potential well,
which cannot be appreciated directly from the plots.
(ii) Furthermore, to compare data and simulations, we use an advanced 
statistical discriminator (Peacock 1982, Fasano \& Franceschini 1987),
which allows us to find quantitative estimates of the probability that
the same process gives rise to observational and simulated statistical
outputs. The discrepancy we find
is however directly visible in suitable evolutionary plots and can be also 
appreciated using a simpler statistical test.

The plan of the paper is as follows.
In Sec. 2 power ratios are defined and their evolution is discussed.
In Sec. 3 we give some details about the model and simulation procedure 
followed to define clusters and study their 
evolution. Power ratios are computed in Sec. 4 where we also perform
the comparison between data and simulations.
Results are discussed in Sec. 5, where some general conclusions are also 
drawn.
Appendix A includes a brief outline of the PF2 statistical test, and
some discussion of the reasons why its extension to three dimensions 
is premature. 

\section{Power ratios: definition and evolution}

In this section we shall report results of Tsai \& Buote (1996) and Buote \& 
Xu (1997), which are the starting point of our arguments.
The steps to work out the power ratios from a model cluster are 
the following ones: (i) We assume that the X--ray emission power
is $\propto \rho_b^2 ({\bf r})$ and project it
on a random plane $\pi_r$, yielding a surface brightness $\Sigma (x,y)$
(apart of a proportionality factor).
(ii) We find the projected mass center (or {\sl centroid}) $O$ for 
$\Sigma (x,y)$; in respect to $O$, the surface brightness can be expressed 
as $\Sigma (R,\varphi)$; (iii) we use such $\Sigma$ to generate a 
pseudo--potential $\Phi(R,\varphi)$, by solving the Poisson equation:
\begin{equation}
\label{Poisson}
\nabla^2 \Phi = \Sigma (R,\varphi) 
\end{equation}
(again, a constant factor, in front of the r.h.s. -- $e.g.$, the
gravitational constant $G$, -- would turn out to be irrelevant).
(iv) We expand $\Phi$ in multipoles; from the coefficients
of such expansion, the power ratios {\apim}$(R)$ are obtained. 
More in detail:
\begin{equation}
\label{prdef}
\Pi^{(m)} (R) = \log_{10} (P_m/P_0)
\end{equation}
where
\begin{equation}
\label{mpdef}
P_m (R) = {1 \over 2 m^2} (\alpha_m^2 + \beta_m^2) ~,
P_0 = [ \alpha_0 \ln(R/{\rm kpc}) ]^2
\end{equation}
and
\begin{equation}
\label{int}
\alpha_m = \int_0^1 { dx x^{m+1} \int_0^{2\pi} { d\varphi [ \Sigma (xR,
\varphi) R^2] \cos(m\varphi) }}
\end{equation}
\begin{equation}
\beta_m = \int_0^1 { dx x^{m+1} \int_0^{2\pi} { d\varphi [ \Sigma
(xR,\varphi) R^2] \sin(m\varphi)}}
\end{equation}
Henceforth, the constants which were unspecified cancel out, when the
ratio of $P_m$'s is considered. Moreover,
owing to the definition of the centroid $O$, $P_1$ vanishes and $\Pi^{(1)}$
should then be disregarded. As we restricted the analysis to $m 
\leq 4$, which account for \sub on scales not much smaller than $R$ itself,
there are only 3 significant \apim ($m=2,3,4$).

Therefore, during its evolution, a cluster moves along a line
({\sl evolutionary track}) in the 3--dimensional space
spanned by such \apim's, whose 
2--dimensional projections will be considered below. Starting from
a configuration away from the origin, corresponding to a large amount
of internal structure, a cluster evolves towards isotropization
and homogeneization. This motion, however, does not occur with a steady trend,
as sudden bursts of structure appear when further matter lumps
approach the cluster potential well, to be absorbed by it.
However, the evolutionary track eventually approaches the origin,
and this can be more easily appreciated by averaging over the
contributions of several clusters.

Actual data, of course, do not show the motion of a single cluster
along the evolutionary track. Different clusters, however, lie at
different redshifts and, in average, can be expected to
describe a succession of evolutionary moments. Power ratios for
model clusters are to be set in the 3--dimensional space
spanned by \apim's, taking each model cluster at redshifts distributed
as for data clusters.

The data set is the same of Buote $\&$ Tsai (1996).
Among X--ray cluster images, taken with $ROSAT$ PSPC (Pfeffermann \etal 1987),
and such that the PSPC central ring contains a cluster portion whose
radius exceeds 400$\, h^{-1}$kpc, they took those contained in 
the HEASARC--legacy database and belonging to the Ebeling
(1993) or Edge \etal (1990) samples. Out of the 59 objects
selected in this way, they could estimate \apim for 44 of them
at $R = 0.8\, h^{-1}$Mpc and for 27 of them at  $R = 1.2\, h^{-1}$Mpc.
PSPC data give the X--ray surface brightness $\Sigma_X (R,\varphi)$,
which is to be used in the same way as the $\Sigma$'s obtained from models,
to work out \apim for $R=0.4,\, 0.8\, ,1.2\, h^{-1}$Mpc.

In fig.~1 we give the redshift distribution of data clusters,
for the three values of $R$. As we shall discuss further in the next section,
we have 40 simulated clusters to deal with. 
\begin{figure}
\vfill
\centerline{\mbox{\epsfysize=7.0truecm\epsffile{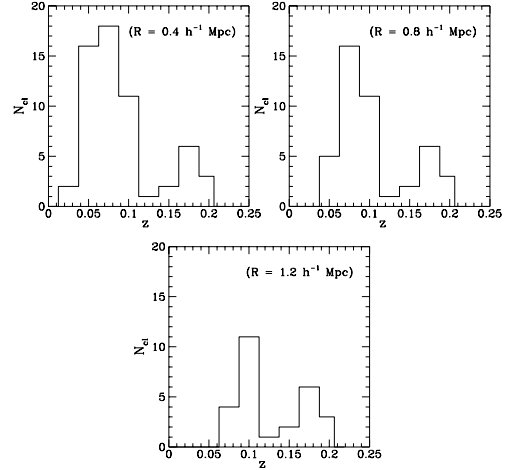}}}
\caption{\it Redshift distribution of data clusters for 
$R=0.4,0.8,1.2 h^{-1} {\rm Mpc}$ }
\label{fig:redhis}
\end{figure}
We take each of them, only at
a single $z$ value, producing 50 $z$ distributions similar to
real data. In order to work out power ratios, we need to treat
each cluster with reference to a given plane: 50 random planes
are selected for each cluster and one of them is
considered for each $z$ sequence. 

\section{The simulations}

The main contribution of this work is given by the simulations we run,
to work out \apim, and that will be however used for several other
aims. We started from a box with a side $L = 200\, h^{-1}$Mpc,
where we evolved a CDM model with $h = 0.5$ and $\Omega_b = 0.06$.
The initial spectrum was normalized in order that the cumulative
cluster number density $n(>M)$, for $M = 4.2\, h^{-1}10^{14} M_\odot$,
yields $N_{cl} = n(>M) L^3 = 32$, according to data on cluster
abundance (White \etal 1993, Biviano \etal 1993, Eke \etal 1996). 
This corresponds to $\sigma_8 = 0.52$. The simulation was run from 
$z_{in} = 6.3$ to $z=0$ using a P3M code with 10$^6$ particles 
(mass $m_p=1.78 \cdot 10^{13} M_\odot$).
The choice of such $z_{in}$ value is dictated by the opportunity of restarting 
from the same conditions when hydrodynamic simulations are then performed.
The PM part of the code makes use of 256$^3$ cells and a smoothing radius of
520 kpc was taken for gravitational forces.

At $z=0$ clusters were then identified using a FoF algorithm; we considered
as {\sl friends}, particles at a distance $< 0.2 \lambda$ ($\lambda$ is
the mean particle separation). The 40 most massive clusters were then
located; the lightest one is made by 34 particles and
has a mass $M_m \simeq 6 \cdot 10^{14} M_\odot$.

For each of these clusters an hydrodynamic simulation was then
performed, using a TREESPH code. 
The hydrodynamical part of the code is based on the SPH method
(Hernquist $\&$ Katz 1989). The gravitational forces are solved with
a hierarchical tree algorithm and this part of the code is based
on the public treecode of L. Hernquist (Hernquist 1987).
In order to run the hdyrodynamical simulation we located the cluster 
center at $z=0$ and found all particles within a radius $r_{200}$, 
where the cluster density is $\simeq 200\, \langle \rho \rangle$ 
(background density). Such particles were traced back to $z_{in}$, 
in the original simulation cube, and a smaller cube of side $L_c$,
enclosing all of them, was then located at the cluster center. We found the 
approximate scaling $L_c \simeq 45\, {\rm Mpc} \times (r_{200}/3.43{\rm Mpc})$.

A high resolution lattice of $22^3$ particles was set in the smaller cube 
and their position was then perturbed, using the same initial conditions
of the cosmological simulations, implemented by additional waves
to sample the increased Nyquist frequency. This technique is similar to the 
one adopted by Katz \& White (1993), Navarro, Frenk \& White (1995).
We used two grids, for baryon and CDM particles, whose masses
$m_b/m_{CDM} = \Omega_b/(1-\Omega_b)$ .

The cube of side $L_c$ is then set inside a greater cube of side $2L_c$,
with parallel sides and equal center. Such greater cube is initially
filled by $22^3$ CDM particles, whose spacing is double in respect
of inner cube particles, and whose mass is $8\, m_{CDM}(1+\Omega_b)$. Then
particles falling in the smaller inner cube are eliminated.
Heavier particle positions are also perturbed, according to the
initial conditions of the cosmological simulation.
Finally, we use all particles inside a sphere, of radius $L_c$ and with the
same center as cubes, for the high resolution TREESPH simulation.

Since we take a constant number of grid particles, their masses
vary according to $L_c$; for the most massive cluster we met
($M_{200} \simeq 10^{15} M_{\odot}$), it was $m_{gas}  = 3.4~10^{10} 
M_{\odot}$. 
 The softening parameter for gravitational forces ($ \varepsilon$)
 was set to $\varepsilon_g=80~kpc$ for gas particles and scales 
as  $\propto m^{1/3}$.
We use a tolerance parameter $\theta=0.7$, without quadrupole
corrections.  Viscosity was set as in Hernquist $\&$ Katz (1989) taking 
their parameters $\alpha = 1$ and $\beta = 2$. 

The minimum timestep was $3~10^6 yr $ for gas particles and twice as much 
for CDM. In average each cluster required $\sim 6\, $hours of CPU time to be 
evolved from $z_{in}$ to $z=0$ on a RISK 6000 elaborator.
Outputs from TREESPH simulations were preserved at various redshifts.
Among the smaller ones, we used $z_i = $0.15, 0.10, 0.049 and 0
($i=1,..,4$). 

\begin{figure}
\vfill
\centerline{\mbox{\epsfysize=7.0truecm\epsffile{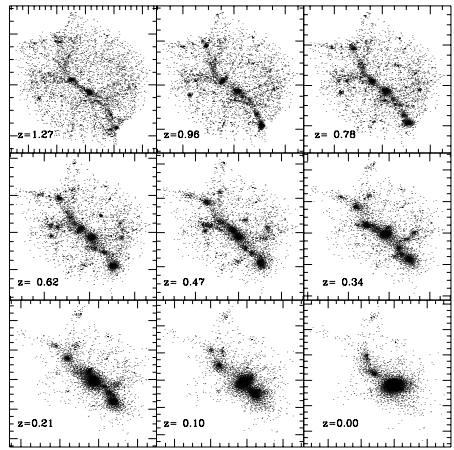}}}
\caption{\it Density projections of a simulated cluster at several redishift.} 
\label{fig:pw34}
\end{figure}

In fig.~2 we give a typical example of projected $\rho_b$ outputs, at a
set of redshifts $\leq 1.27$. Already from such figure, it is
clear that CDM clusters are rich of substructures and characterized by
a fast substructure evolution. In fig.~3 we show CDM
and baryon distributions at $z=0$ for a typical cluster.
Such figure shows that the baryon distribution is significantly
smoothed out in respect to CDM. This obviously reflect on \apim's,
although we shall report no quantitative details on this point here.

\begin{figure}
\vfill
\centerline{\mbox{\epsfysize=7.0truecm\epsffile{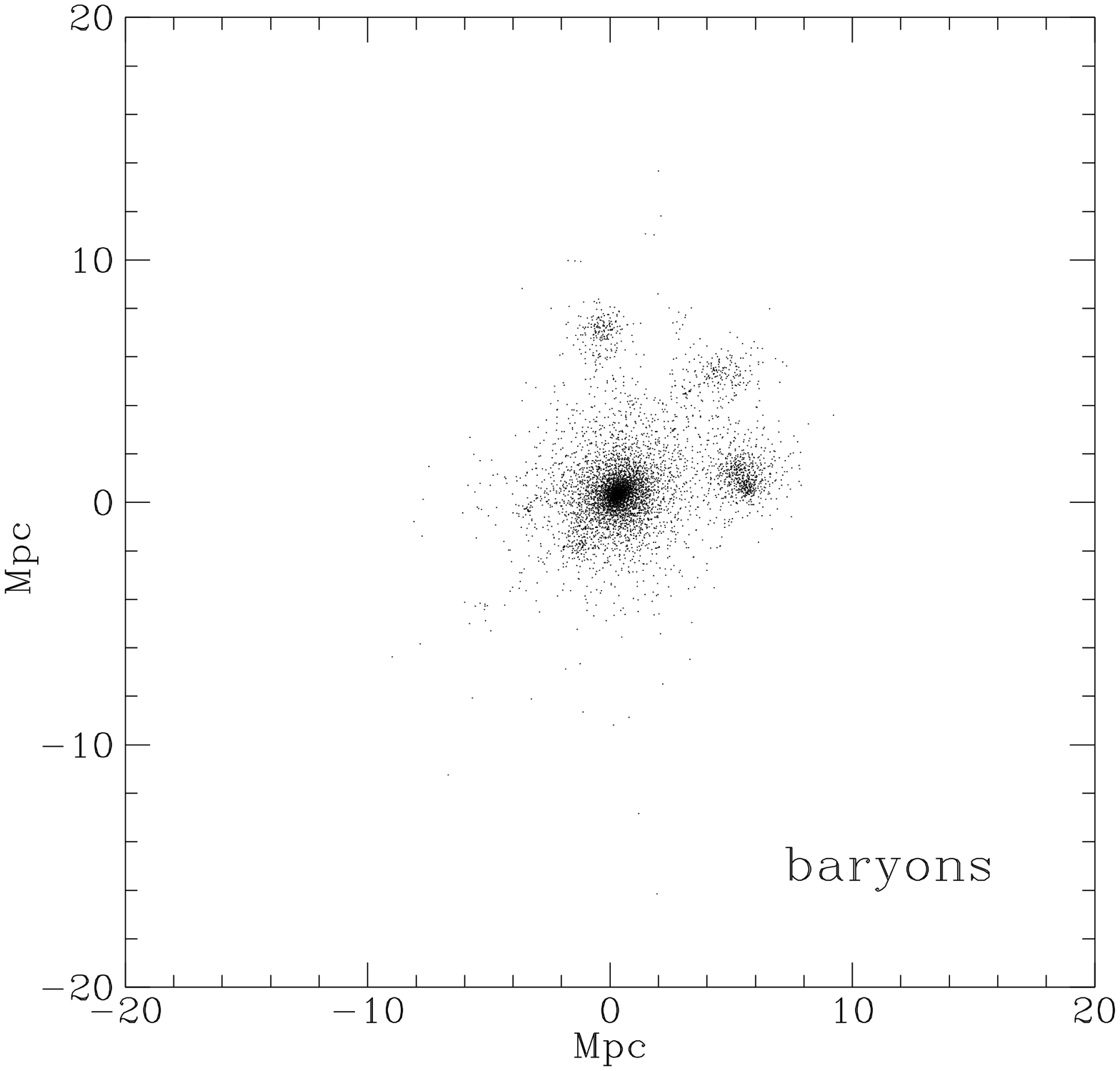}}}
\centerline{\mbox{\epsfysize=7.0truecm\epsffile{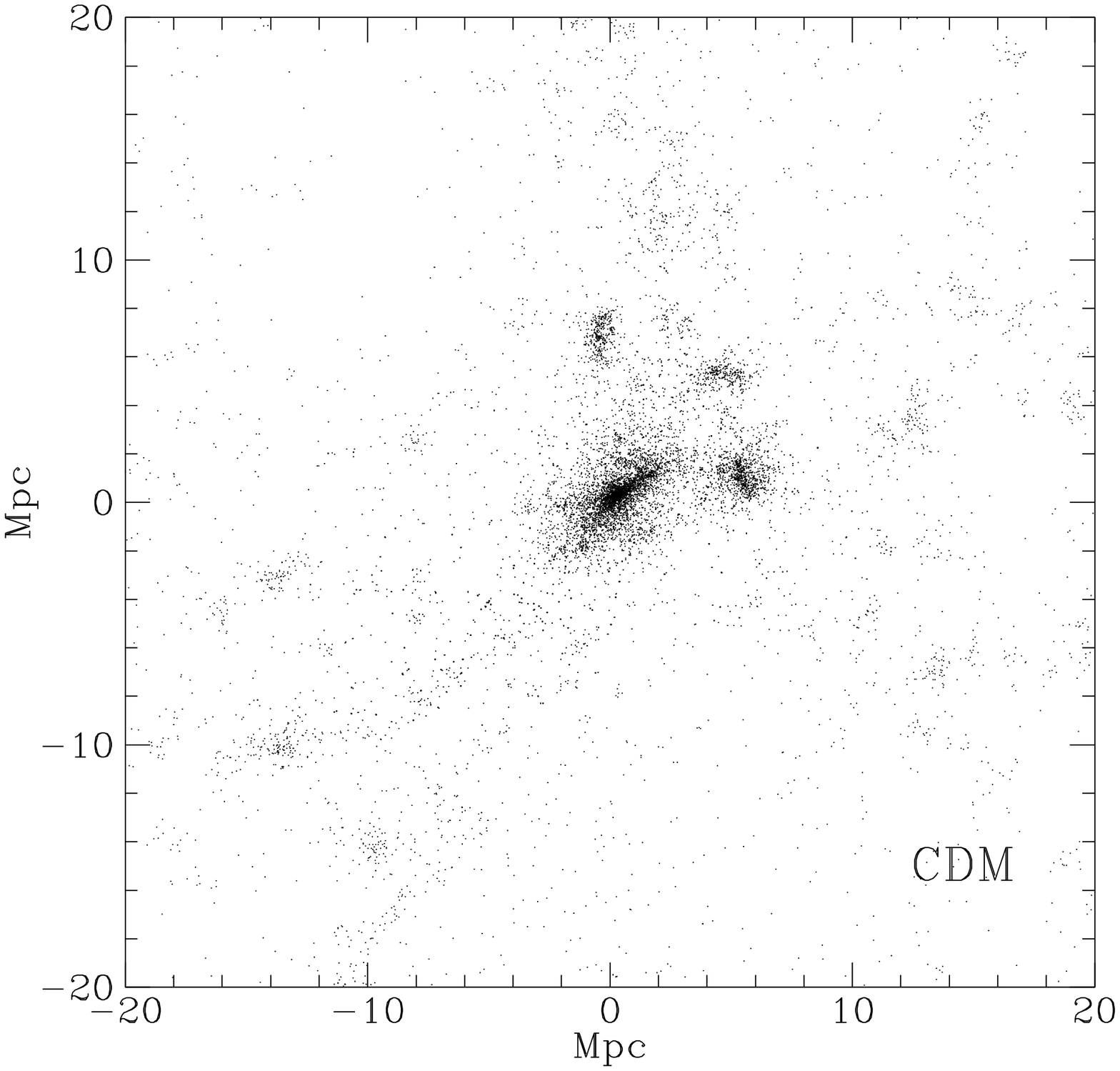}}}
\caption{\it  Baryon  and dark matter density projections   of a simulated
cluster}
\label{fig:gasdm}
\end{figure}

\section{Cluster evolution from power ratios behaviour}

Clusters are expected to evolve by accreting lumps of baryons and CDM
onto a deep potential well. As time elapses, nearby lumps are gradually
absorbed and cluster substructures tend to vanish. Such behaviour is
clearly visible also in fig.~2.
In fig.~4, instead, we give 9 plots for the space containing the
evolutionary tracks projected on 3 planes and at three different $R$'s.
Cross indicate the location of data clusters. Filled dots refer to
a single selection of simulated clusters. The trend visible in
this selection of simulated clusters is however a general feature:
a linear regression gives place to the dotted straight line
reported in the plots. Although real data do not show such a marked
trend, a linear regression was performed also on them. The continuous
lines show the results of such linear fits to data.

\begin{table}
\begin{center}
\caption{Angular coefficients and standard deviations for linear fits.
Indeces $_d$ $_s$ refers to data and simulations. In parenthesis we
give also simulation errors from 20 bootstrap resamplings.}
\begin{tabular}{c| c |c |c |c |c }
\hline \hline
$R\, h/{\rm Mpc}$&& $m_{d}$ & $\Delta m_{d}$& 
$m_{s}$& $\Delta m_{s}$
\\ \hline
   & $\Pi_2$ vs. $\Pi_3$ & 0.29 & 0.19 & 0.64 & 0.18 (0.036)
\\
0.4& $\Pi_2$ vs. $\Pi_4$ & 0.52 & 0.14 & 0.85 & 0.12 (0.026)
\\
   & $\Pi_3$ vs. $\Pi_4$ & 0.38 & 0.10 & 0.61 & 0.11 (0.011)
\\ \hline
   & $\Pi_2$ vs. $\Pi_3$ & 0.51 & 0.15 & 0.87 & 0.15 (0.020)
\\
0.8& $\Pi_2$ vs. $\Pi_4$ & 0.91 & 0.14 & 1.03 & 0.12 (0.010)
\\
   & $\Pi_3$ vs. $\Pi_4$ & 0.71 & 0.14 & 0.92 & 0.07 (0.004)
\\ \hline
   & $\Pi_2$ vs. $\Pi_3$ & 0.43 & 0.16 & 0.87 & 0.10 (0.008)
\\
1.2& $\Pi_2$ vs. $\Pi_4$ & 0.18 & 0.20 & 0.97 & 0.11 (0.008)
\\
   & $\Pi_3$ vs. $\Pi_4$ & 0.51 & 0.21 & 1.03 & 0.07 (0.009)
\\ \hline\hline
\end{tabular}
\end{center}
\end{table}
\begin{table}
\begin{center}
\caption{Kolmogorov--Smirnov test on angular coefficients for the sequence 
yielding  fig. 4 and averaged on 50 redshifts sets.}
\begin{tabular}{c| c  |c  |c }
\hline \hline
$R\, h/{\rm Mpc} $&&  KS--test& KS--test (av.)
\\ \hline
   & $\Pi_2$ vs. $\Pi_3$ & 1.61E-22 & 3.21E-23 
\\
0.4& $\Pi_2$ vs. $\Pi_4$ & 6.43E-32 & 1.29E-32
\\
   & $\Pi_3$ vs. $\Pi_4$ & 3.79E-29 & 7.60E-30
\\ \hline
   & $\Pi_2$ vs. $\Pi_3$ & 1.48E-22 & 1.62E-08
\\
0.8& $\Pi_2$ vs. $\Pi_4$ & 5.75E-05 & 0.12E+00
\\
   & $\Pi_3$ vs. $\Pi_4$ & 1.72E-19 & 1.65E-16
\\ \hline
   & $\Pi_2$ vs. $\Pi_3$ & 1.19E-20 & 3.41E-16
\\
1.2& $\Pi_2$ vs. $\Pi_4$ & 6.52E-24 & 6.24E-24
\\
   & $\Pi_3$ vs. $\Pi_4$ & 4.45E-22 & 1.07E-18
\\ \hline\hline
\end{tabular}
\end{center}
\end{table}	
Fig.~4 shows that model lines are steeper than data lines {\sl
in all cases}. This behaviour is true for all model sets we explored
and indicates a faster evolution of model clusters in respect to the
real ones. In Table 1, we report the values of the angular
coefficients $m$ of the two straight lines of each plot, together with 
$1\, \sigma$ errors $\Delta m$, obtained from the linear regression. 

Another way of testing the stability of $m$ amounts to performing a
bootstrap--resampling of the simulation points. In fact,
the 2-dimensional point distributions are rather irregular
and a bootstrap approach tests the reliability
of the linearity assumption, together with the actual $m$ values.

Bootstrap resampling was performed selecting clusters with the
correct $z$ distribution, but allowing data repetition. 
All resamplings confirmed that model fits are steeper than data.
Furthermore, bootstrap errors are systematically smaller than
errors from linear regressions. Average errors
from 20 bootstrap resamplings are reported in parenthesis
in Table 1, aside of those from linear fits.

In Table 2 we report the
the results of a Kolmogorov--Smirnov test applied to the line slopes
of the single
case considered in Fig. 4, and the values obtained by averaging
on the set of 50 realizations.
The figures are the probabilities that the different slopes we find in
data and models can be due to the same process, taking into
account the number of data points in use.
\begin{figure}
\vfill
\centerline{\mbox{\epsfysize=7.0truecm\epsffile{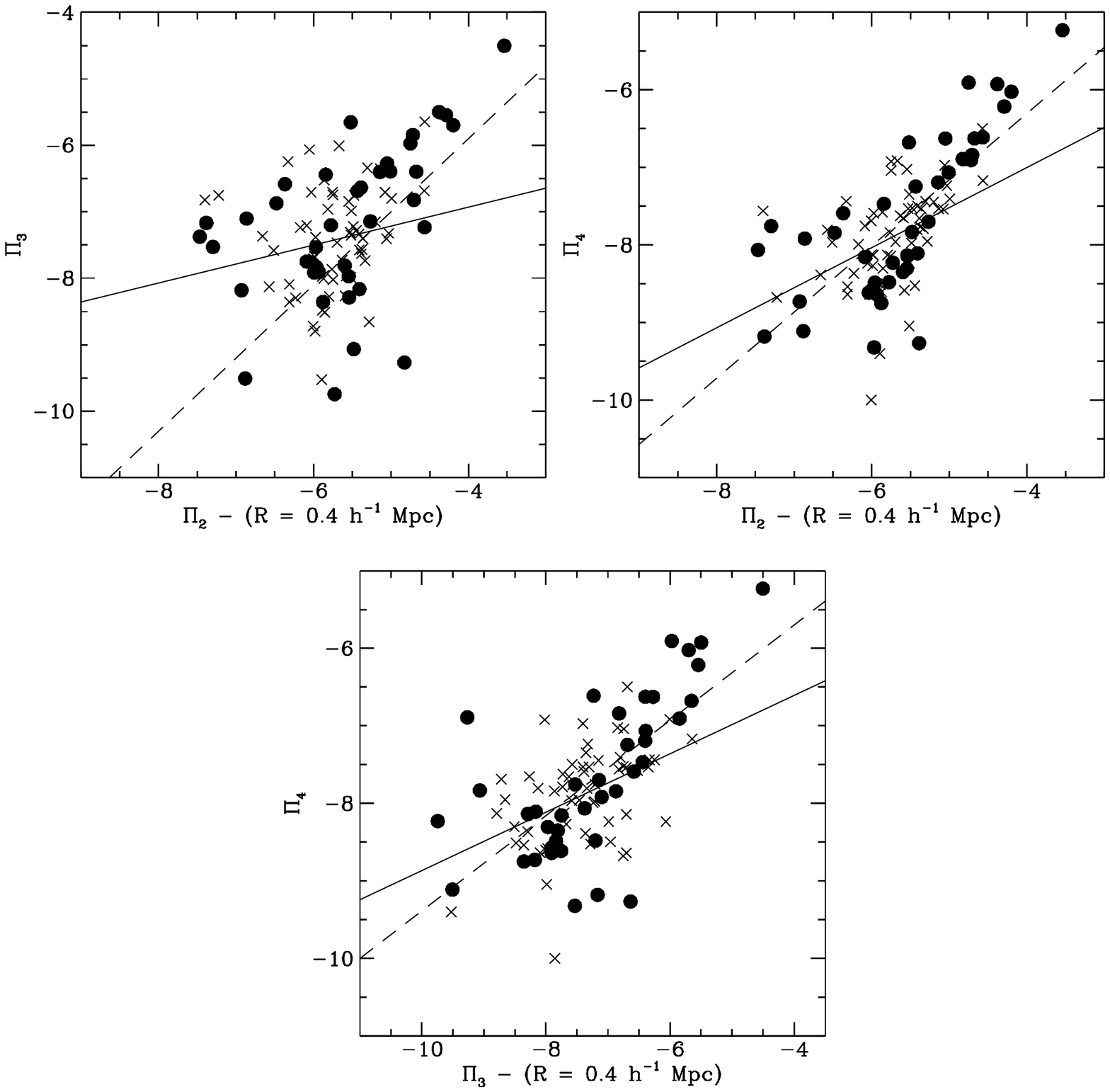}}}
\centerline{\mbox{\epsfysize=7.0truecm\epsffile{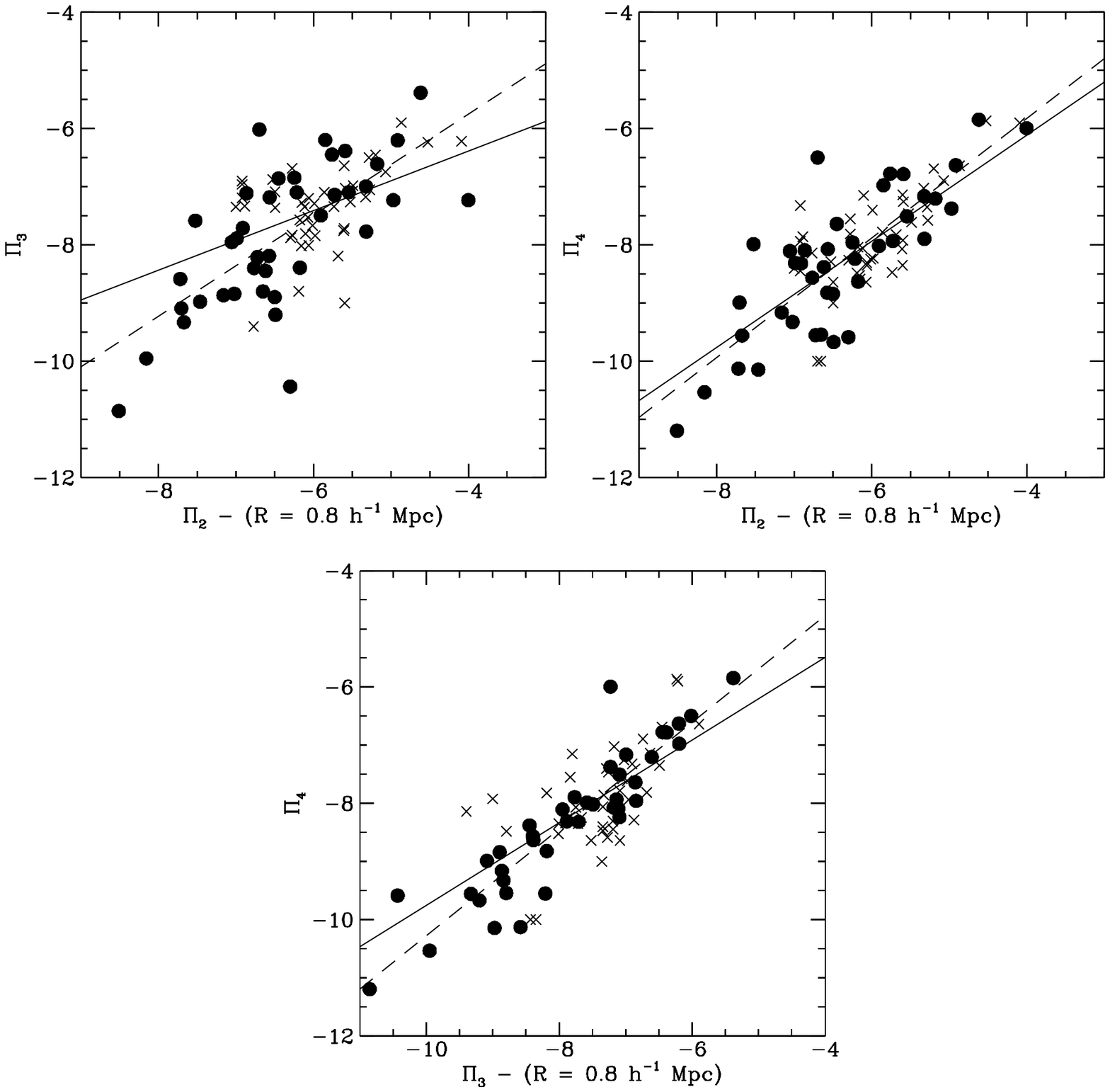}}}
\centerline{\mbox{\epsfysize=7.0truecm\epsffile{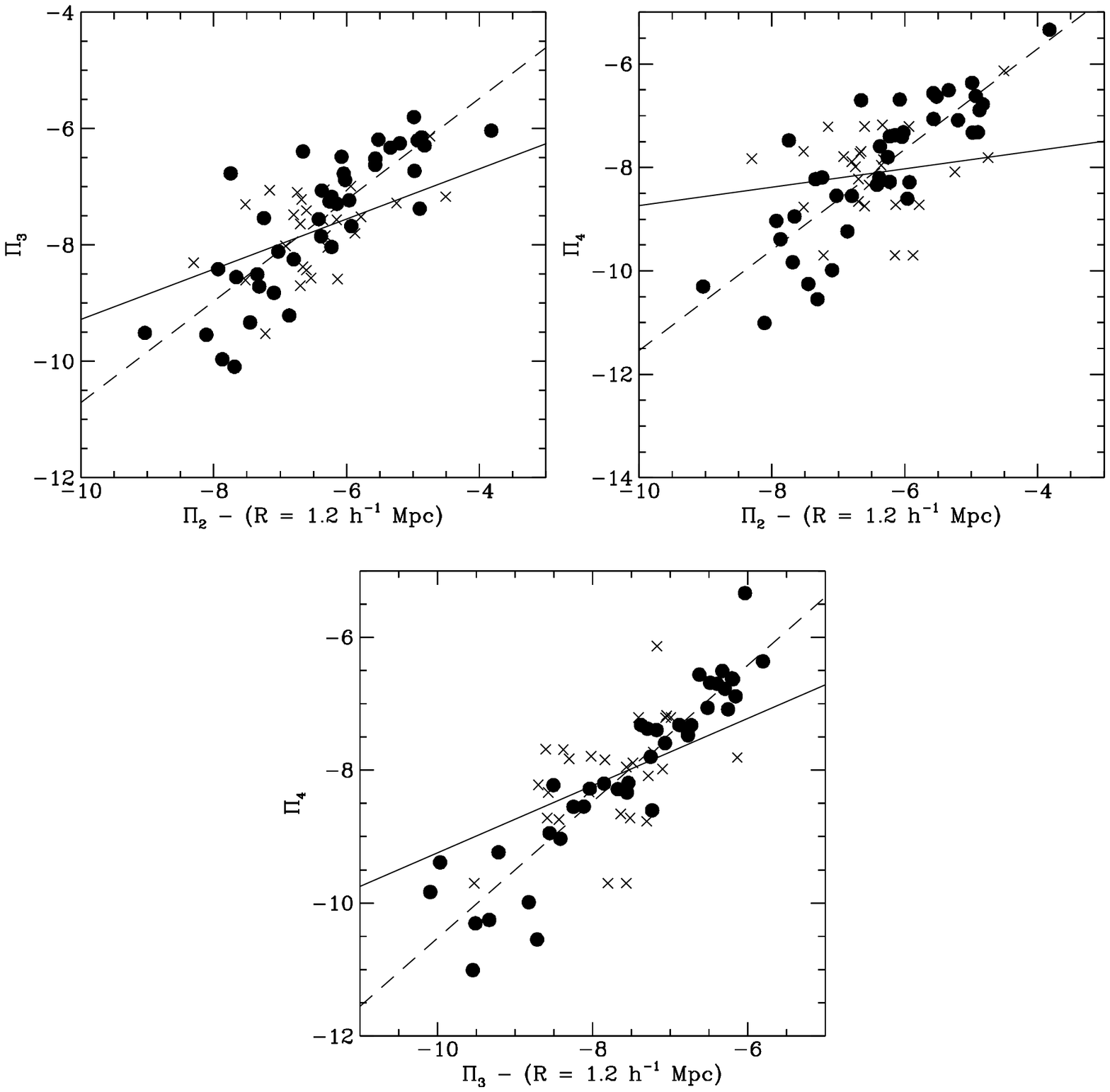}}}
\caption{\it Power ratios and evolutionary tracks for data and simulated
clusters at three aperture radii.
Crosses and continuous lines refer to the observational sample, filled dots and
dashed lines refer to one of the artificial sequences of 40 clusters,
with similar $z$ distribution. Plots obtained from
all 50 cluster sequences are strictly analogous.}
\label{fig:pw}
\end{figure}
This test reproduces the visual feeling that the two straight lines,
coming from data and simulations, have a systematically different slope.
(Using bootstrap errors as $\Delta m$, besides of being scarcely significant,
would lead to even smaller probabilities.)
This test, however, is somehow misleading and actual probabilities
are substantially greater than so.
\begin{table}
\begin{center}
\caption{PF2-- test on $\Pi_m$  for the sequence 
yielding  fig. 4 and averaged on 50 redshifts sets.}
\begin{tabular}{c| c  |c |c }
\hline \hline
$R \, h/{\rm Mpc}$&&  PF2--test & PF2--test (av.)
\\ \hline
   & $\Pi_2$ vs. $\Pi_3$ & 0.129 & 0.057
\\
0.4& $\Pi_2$ vs. $\Pi_4$ & 0.045 & 0.052
\\
   & $\Pi_3$ vs. $\Pi_4$ & 0.073 & 0.079
\\ \hline
   & $\Pi_2$ vs. $\Pi_3$ & 0.017 & 0.076
\\
0.8& $\Pi_2$ vs. $\Pi_4$ & 0.038 & 0.087
\\
   & $\Pi_3$ vs. $\Pi_4$ & 0.031 & 0.052
\\ \hline
   & $\Pi_2$ vs. $\Pi_3$ & 0.038 & 0.053
\\
1.2& $\Pi_2$ vs. $\Pi_4$ & 0.050 & 0.034
\\
   & $\Pi_3$ vs. $\Pi_4$ & 0.016 & 0.054
\\ \hline\hline
\end{tabular}
\end{center}
\end{table}
The extra input that we should eliminate, to obtain
the probability that the observational data set
is fit by artificial data sets, is the requirement of linear
behaviour. This can be done by applying the
generalization of the Kolmogorov--Smirnov test to 2--dimensional
data sets, proposed by Peacock (1982) and widely tested by
Fasano $\&$ Franceschini (1989). We shall refer to this
estimate as PF2 test and more details on it will be given
in Appendix A. In Table 3 we give 
the values of the single case of Fig. 4 and 
the  averaged values of the
PF2--probabilities that the observational
data set and each artificial data set are realizations of the
same process, for each $\Pi_m$
pair and each $R$ value. Such average PF2--probabilities are not so small
as those reported in Table 2 and essentially correspond to a
$\sim 2\, \sigma$ discrepancy. However, even this sophisticated
test is unable to work out a final value which takes
into account all the above partial results. 
There can be however
little doubt left that such a sequence of values, where
a faster evolution in model in respect to data is systematic,
could hardly be due
to observational sample biases or to disregarding peculiar physical
effects in the simulations, and is more likely to indicate
that the model is not adept to reproduce real data.

\section{Conclusions}

It has been known for several years that COBE normalized CDM yields
a cluster abundance exceeding observation by a factor $\sim 20$.
In order to obtain a fair number of clusters, the quadrupole data
has to be lowered down to $\sim 9\, \mu$K, at more than 3$\, \sigma$'s
from the observational value. However, CDM models are still currently
in use as reference models; in fact, 0CDM models, mixed models or 
$\Lambda$ models, used to obtain fair fits of
observational data on various scales, involve more parameters.

According to our results, however, even though CDM is normalized to
yield a fair number of clusters, we cannot expect their morphologies
to be adequately approached by the model. This output is related to
the slope of the transfered spectrum, as already outlined in Sec.~1,
and as is expressed by the values of the {\sl extra power} parameter
$\Gamma$. It is however important and fairly unexpected that the
effects of such different slope are already visible in the 
average evolutionary tracks, for a fairly narrow redshift range ($0<z<0.2$)
as the one we had in observational data.

In a forthcoming paper, based on work in progress, we shall 
extend this analysis to other
cosmological models, whose transfered spectrum has a $\Gamma$\
value consistent with observations. This is important, in order to
test how sensitive the power ratios $\Pi_m$ 
are to the $\Gamma$ value and to check
whether this test depends only on $\Gamma$ or, $e.g.$, also on the
$substance$ of the model considered.

\section*{Acknowledgments}
It is a pleasure to thank David Buote for his useful comments. 
One of us (S.G.) wishes to thank SISSA for its hospitality during 
the preparation of this work.

\vskip 0.5truecm
\noindent
{\bf APPENDIX A -- The PF2 test}

\vskip 0.3truecm
\noindent
As is known, the Kolmogorov--Smirnov (KS) test is used to evaluate
the probability that 2 distributions on a single variable $x$ are produced
by the same process. If $N$ events are located at $x_i$ ($i=1,..,N$),
let $S_{N}(x)$ give the fraction of events located at $x_i < x$. Clearly
$S_{N}$ is discontinuous in each $x_i$, where it increases by $1/N$.
Let two sets of $N_1$ and $N_2$ events yield $S_{N_1}$ and $S_{N_2}$; the KS 
statistics is then defined as the maximum distance between them:
$$
D = max_{\{any~x\}} |S_{N_1} (x) - S_{N_2} (x)|
\eqno (a1)
$$
and it can be shown that the probability that the two data sets are
produced by the same process can be estimated, by using $D$, as follows:
Let be 
$$
1/\nu^2 = 1/N_1 + 1/N_2
\eqno (a2)
$$
and $u=D(\nu + 0.12 + 0.11/\nu)$. Then such probability reads:
$$
KS(u) = 2\sum_{r=1}^\infty [(-1)^r / \exp(2r^2u^2)] ~.
\eqno (a3)
$$
Various numerical routines exist which allow to evaluate such function
with any desired approximation (see, $e.g.$, Press \etal 1992).

While for one--dimensional distributions the meaning and the building of $D$
are simple (it tells us the maximum difference of integrated distributions),
in the case of two--dimensional distributions the events are not ordered
and there is no immediate way of telling where such difference is maximum.
A further problem is set by the degree of correlation amongst data.
A limiting situation occurs if they locate along a straight line on a
plane, and then the KS test would be suitable again.

The problem was considered in detail by Peacock (1982) and Fasano $\&$
Franceschini (1987), whose approach (PF2) we shall briefly report here.
If $N$ data are located on the 2--dimensional plane $x,y$ and have average 
$\bar x$,$\bar y$, let us first define $\xi_i = x_i - \bar x$
and $\eta_i = y_i - \bar y$. Their correlation coefficient is then:
$$
r = (\Sigma_i \xi_i \eta_i) / (\sqrt{\Sigma_i \xi_i^2} 
\sqrt{\Sigma_i \eta_i^2})
\eqno (a4)
$$
(notice that, for $r=1$, the points would actually lie on a straight line).
Then, if we have two sets of $N_1$ and $N_2$ data, with correlation 
coefficients $r_1$ and $r_2$, defined according to eq.~(a4), we define
$$
{\tilde r} = {1 \over 2}\sqrt{r_1^2 + r_2^2}
\eqno (a5)
$$
as square average correlation. 

The basic point, however, amounts to define a 4--value extension of the 
function $S_N$, for each set of $N$ data: the 4 values are the fractions
of data in each of the four {\sl quadrants} set by the point
$x,y$ where $S_N$ is evaluated. The PF2
statistics $D$ will be the maximum difference between $S_{N_1}$ and $S_{N_2}$ 
values, ranging both over points and quadrants. Using such $D$, the
square average correlation $r$, and $\nu$, defined according to eq.~(a2),
we then define
$$
u = \nu^2 D/[\nu + \sqrt{1-{\tilde r}^2}(0.25\nu - 0.75)]
\eqno (a6)
$$
and, using eq.~(a3), we obtain the required probability.

It ought to be outlined that there are some restrictions to the
significance of these probability estimators. In general, they become
safer as $\nu$ increases. However, in the one--dimensional case,
the estimate is good for $\nu \simgt 10$--20 (corresponding to $N_1 \sim
N_2 \sim 8$). In the two--dimensional case, the restriction is already
much stronger, as it must be $\nu \simgt 100$--400 (corresponding to
$N_1 \sim N_2 \sim 25$), and the further condition that the probability
found is $\simlt 20\, \%$ is to be added.

Our sets of data and simulation points are therefore just adequate
to meet the above conditions. This is important also in view of
a step forward that could be considered: treating directly the
3--dimensional distributions in the parameter space spanned
by $\Pi^{(m)}$ ($m=2,3,4$), instead of their two--dimensional projections.
An algorithm extending the above test to 3--dimensions is a straightforward
generalization; the range of validity of such algorithm, instead, should
be accurately tested. A large portion of the work of Fasano and Franceschini
(1987) amounted to perform such checks, for the two--dimensional case.
However, the trend found beween 1 and 2 dimensions, seems to indicate
that, in three dimensions, we need 2 sets of data containing $\sim 10^2$
points. As far as simulations are concerned, this requires just
doubling our simulation set. Data clusters, instead, are signicantly
below such limit.

\bsp

\end{document}